\documentclass[12pt]{article}

\usepackage{epsfig,times}

\def  \Bcet   {$B^-_c \to \eta' \ell^- \bar{\nu}~$}

\def  \bcen  {\begin{center}}
\def  \ecen  {\end{center}}

\def  \be    {\begin{equation}}
\def  \ee    {\end{equation}}
\def  \bea  {\begin{eqnarray}}
\def  \eea  {\end{eqnarray}}
\def  \bfig  {\begin{figure}}
\def  \efig  {\end{figure}}

\def  \ra    {\rightarrow}

\begin{document}

\begin{titlepage}

\renewcommand{\thefootnote}{\fnsymbol{footnote}}

\begin{flushright}
{\tt hep-ph/0509143}
\end{flushright}

\vskip 1.2cm

\bcen {\Large \bf \boldmath  \Bcet decay and lepton
polarization asymmetry } \\

\vskip 1.8cm

{\sc \sffamily  V.
Bashiry$^1$\footnote{bashiry@newton.physics.metu.edu.tr}} \vskip
.8cm
$^1${\sl ENGINEERING FACULTY, \\
CYPRUS INTERNATIONAL UNIVERSITY,\\ Via Mersin 10, Turkey.}
 \\[2ex]
 \ecen

\vskip 1.6cm
\begin{abstract}
\noindent In this paper we study the lepton polarization asymmetry
for the simileptonic OZI-forbidden annihilation \Bcet decay where
$l=\mu,\tau$. Our results show that the branching ratio turn out to
be of order $10^{-4}$. Beside, we find that longitudinal,
transversal and normal components of lepton polarizations can be
measured for both $\mu$ and $\tau$ decay modes in the future
experiments at the LHC.

\vspace{2cm} PACS numbers 14.40, Nd 13.20, He 13.60-r, 11.30.Er
\end{abstract}
\newpage
\setcounter{page}{1}

\setcounter{footnote}{0}
\renewcommand{\thefootnote}{\arabic{footnote}}
\section{Introduction}
The $B_c$ meson was first observed in  CDF\cite{CDF,CDF1} detector
at the Fermilab Tevatron in 1.8 TeV $p\bar p$ collisions. It was
measured to have a mass $M_{B_c}=6.40\pm0.39\pm0.13 GeV$ and
lifetime $\tau_{B_c}=0.46^{+0.18}_{-0.16}\pm0.03ps$, which agree
with the theoretical predictions\cite{production, beneke}. Its mass
and the spectrum of the binding system can be computed by potential
model\cite{potential,prod}, PNRQCD\cite{prod,pnrqcd} and lattice
QCD\cite{latt} etc. The results are in the region, $m_{B_c}\simeq
6.2\sim 6.4$GeV. Its lifetime was estimated in terms of the
effective theory of weak interaction and by applying the effective
Lagrangian to the inclusive processes of $B_c$
decays\cite{prod,dec,bigi,changli}. According to the estimates, the
lifetime is $\tau_{B_c}\simeq 0.4$ps, a typical one for weak
interaction via virtual $W$ boson. Further detailed experimental
studies can be performed at B factories ( KEK , SLAC) and CERN large
Hadron Collider(LHC). Especially, at LHC with the luminosity ${\cal
L}=10^{34}cm^{-2}s^{-1}$ and $\sqrt{s}=14\rm TeV$, the number of
$B_c^{\pm}$ events is expected to be about $10^{8}\sim10^{10}$ per
year \cite{LHC}, so there seems to exist a real possibility to study
not only some $B_c$ rare decays, but also CP violation, T violation
and polarization asymmetries. The studies of  CP violation, T
violation and polarization asymmetries are specially interesting
since they can serve as good tools to test the predictions of SM or
to reveal the new physics effects beyond the SM.

The study of  $B_c$ meson,  the ground state of the heavy-flavored
binding system $(\bar{c}b)$, the bottom and the charm, constitute a
very rich laboratory since  this meson is also a suitable object for
studying the predictions of QCD. As $B_c$ meson has  many decay
channels because of its sufficiently large mass predictions of QCD
are more reliable. The $B_c$ meson decays provide windows for
reliable determination of the CKM matrix element $V_{cb}$ and can
shed light on new physics beyond the standard model.

In the framework of the SM its decays can occur via three
mechanisms: (1) the c-quark decay with the b-quark being a
spectator, (2) the b-quark decay with the c-quark being a spectator,
(3) b-quark and c-quark annihilation. The first two mechanisms are
expected to contribute about $90\%$ of the total width, and the
remaining $10\%$ is owed to the  annihilation process.

\par
There is another decay mode which does not belong to the three
aforementioned types, and it can only occur via the OZI processes.
As we know that the OZI rule\cite{OZI} plays an important role in
the processes which occur via strong interaction and in general at
the parton level the concerned calculations are carried out in the
framework of the perturbative QCD (PQCD).

The first investigation of the OZI-forbidden annihilation decays
\Bcet  in  QCD was carried out in 1999 \cite{akio}.  In their work,
an effective Lagrangian was adopted to avoid introducing the $B_{c}$
meson wave function, meanwhile they dealt with the light meson by
using an effective $g_{a}^*g_{b}^*\to \eta^{'}$ coupling
\cite{yang}\cite{close}, which was obtained in the NRQM
approximation. The valence quark $q$ and anti-quark $\bar{q}$ in the
light meson were assumed to possess equal momenta and be on their
mass shells, i.e., $p_q=p_{\bar{q}}$ and $p_q^2=m_q^2$. The
branching ratio estimated to be $ Br(B_c \to \eta' \ell
\bar{\nu})=1.6\times 10^{-4}$ for $l=\mu,e$, which is accessible at
CERN LHC.

 In this paper,  we investigate lepton polarization
  asymmetries in semileptonic annihilation decay \Bcet.

The paper is organized as follows. In section 2, we give the details
of the calculation of  the amplitude and polarization asymmetries.
Section 3 is devoted to numerical results and discussions.

\section{Calculations}

The effective Lagrangian responsible for $(b\bar{c})\rightarrow
g^*_a g^*_b l\bar\nu$ decay is \cite{akio} \bea {\cal M} (b\bar c
\ra g^*_a g^*_b l^- \nu )= \frac{G_{F}}{\sqrt{2}}
 V_{cb}g_s^2 Tr[T_a T_b] \bar{v}_c (p_c )
 \left[
       \gamma_{\mu}(1-\gamma_{5} )
  \frac{i}{\slash{\hskip -0.23cm}p_{b} -\slash{\hskip -0.23cm}K -m_b}
  \gamma_{\beta}
  \frac{i}{\slash{\hskip -0.23cm}p_{b} -\slash{\hskip -0.23cm}k_1 -m_b}
  \gamma_{\alpha} \right. \nonumber\\
\left. +\gamma_{\alpha} \frac{i}{\slash{\hskip -0.23cm}k_{1}
-\slash{\hskip -0.23cm}p_{c}-m_c}
  \gamma_{\beta}
\frac{i}{\slash{\hskip -0.23cm}K-\slash{\hskip -0.23cm}p_{c}-m_c}
  \gamma_{\mu}(1-\gamma_{5} )
+\gamma_{\beta} \frac{i}{-\slash{\hskip -0.23cm}p_{c}+\slash{\hskip
-0.23cm}k_{2} -m_c}
  \gamma_{\mu}(1-\gamma_{5} )
  \frac{i}{\slash{\hskip -0.23cm}p_{b}-\slash{\hskip -0.23cm}k_{1} -m_b}
  \gamma_{\alpha}\right] u_{b}(p_{b})
\nonumber \\
\times
  {\bar l}\gamma^{\mu}(1-\gamma_{5})\nu_{l}
+\left(\alpha\leftrightarrow\beta, k_{1}\leftrightarrow k_{2}
\right). \eea Using the Dirac equation and identity for Dirac
matrices, after long but straightforward calculations for an
effective Lagrangian ${\cal L}_{eff}$ we get\cite{akio}: \\
\be {\cal L}_{eff}= \frac{G_{F}}{\sqrt{2}}
 V_{cb}g_{s}^{2} Tr[T_{a}T_{b}]
 \bar{c}\gamma_{\delta}(1-\gamma_{5})
b{\hskip 2mm} {\bar l}\gamma_{\mu}(1-\gamma_{5})\nu_{l}
 {\cal F}^{\delta\mu\alpha\beta}\frac{1}{k^2_1}\frac{1}{k^2_2}
\langle g^*_{a\alpha} g^*_{b\beta} \mid\eta^{\prime}\rangle. \ee
where $K=k_1+k_2$ is the momentum of $\eta'$, ${\cal
F}^{\delta\mu\alpha\beta}$ represents the combination of momenta and
contains loop integrations, which is  explicitly shown in
\cite{akio}.

Having obtained the effective Lagrangian, the total amplitude can be
obtained by sandwiching the ${\cal L}_{eff}$ between annihilated
meson state  $| B^-_c\rangle $ and created non-meson state
$\langle0| $by using the definition \be \langle 0\mid
\bar{c}\gamma_{\mu}(1-\gamma_{5})b\mid B_{c}(P)\rangle=
if_{B_c}P_{\mu} \ee and the $g^*_a g^*_b \rightarrow\eta^{\prime}$
coupling \be \langle g^*_a g^*_b \mid\eta^{\prime}\rangle=g_s^2
\delta_{ab} \frac{A_{\eta^{\prime}} }{k_{1}\cdot
k_{2}}\epsilon_{\alpha\beta m n } k_{1}^m k_{2}^n \ee which has been
widely used in $\eta^{\prime}$ and pseudoscalar productions in heavy
quarkonium decays and in high energy collidors\cite{close}.

Here the
parameter $A_{\eta^{\prime}}$ is understood as a combination of
$SU(3)$ mixing angles and nonperturbative objects,  and can be
extracted from the decay $J/\Psi\rightarrow\eta^{\prime}\gamma$.

Using the definitions mentioned above and performing the loop
integrations via Dimensional Regularization, the amplitude is found
as follows:\be {\cal M}=\frac{G_{F}}{\sqrt{2}}
 V_{cb}g_s^4 Tr[T_a T_b]\delta_{ab}4A_{\eta^{\prime}} if_{B_c}
 \frac{i}{16\pi^2}\left( P_{\mu}f_{1}+K_{\mu}f_{2}\right)
 \bar{l}\gamma^{\mu}(1-\gamma_{5})\nu_{l}
\ee
 Where $P_{\mu}$ and $K_{\mu}$ are the momentums of  $B_c$ and $\eta^{\prime}$, respectively.
 With $f_{1}, f_{2}$ defined by \bea f_{1}&=&-4C_{11}(K,
p_{b}-K, 0, 0, m_{b})+4C_{12}(K, p_{b}-K, 0, 0,m_{b})
\nonumber \\
&&-2C_{11}(\frac{K}{2}, \frac{K}{2}-p_{b}, 0,
       \frac{m_{\eta^{\prime}}}{2}, m_{b})
   -2C_{12}(\frac{K}{2}, \frac{K}{2}-p_{b}, 0,
      \frac{m_{\eta^{\prime}}}{2}, m_{b})
\nonumber \\
&&-4C_{11}(K, p_{c}-K, 0, 0, m_{c})+4C_{12}(K, p_{c}-K, 0, 0,m_{c})
\nonumber \\
&&+2C_{11}(\frac{K}{2}, \frac{K}{2}-p_{c}, 0,
       \frac{m_{\eta^{\prime}}}{2}, m_{c})
   +2C_{12}(\frac{K}{2}, \frac{K}{2}-p_{c}, 0,
      \frac{m_{\eta^{\prime}}}{2}, m_{c})
\nonumber \\
&& +\frac{2m_{b}}{m_{c} }C_{12}(\frac{K}{2}, p_{b}-K, 0,
       \frac{m_{\eta^{\prime}}}{2}, m_{b})
-\frac{2m_{c}}{m_{b} }C_{12}(\frac{K}{2}, p_{c}-K, 0,
       \frac{m_{\eta^{\prime}}}{2}, m_{c})
\nonumber \\
&&
-\frac{2M(m_{b}-m_{c})}{m_{b}m_{c}}\left(
C_{12}(\frac{K}{2}-p_{c}, P-K,
       \frac{m_{\eta^{\prime}}}{2}, m_{c}, m_{b})
\right.  \\
&& \left. -\frac{m_{c}}{M}C_{11}(\frac{K}{2}- p_{c}, P-K,
       \frac{m_{\eta^{\prime}}}{2}, m_{c}, m_{b}) \right), \nonumber
\eea
and
{\small
\bea
&f_{2}&=\frac{-4Mm_{b}}{K^2 -2p_{b}{\cdot}K}
\left(
2C_{11}(K, p_{b}-K, 0, 0, m_{b})-C_{12}(K, p_{b}-K, 0, 0,m_{b})
+C_{11}(\frac{K}{2}, \frac{K}{2}-p_{b}, 0,
       \frac{m_{\eta^{\prime}}}{2}, m_{b})
\right)  \nonumber \\
&&+\frac{4Mm_{c}}{K^2 -2p_{c}{\cdot}K}
\left(
2C_{11}(K, p_{c}-K, 0, 0, m_{c})-C_{12}(K, p_{c}-K, 0, 0,m_{c})
+C_{11}(\frac{K}{2}, \frac{K}{2}-p_{c}, 0,
       \frac{m_{\eta^{\prime}}}{2}, m_{c})
\right)
\nonumber \\
&&+\frac{M}{m_{c}}
\left(
C_{11}(\frac{K}{2}, p_{b}-K,
       \frac{m_{\eta^{\prime}}}{2}, 0, m_{b})
-2C_{12}(\frac{K}{2}, p_{b}-K,
       \frac{m_{\eta^{\prime}}}{2}, 0, m_{b})
+C_{0}(\frac{K}{2}, p_{b}-K,
       \frac{m_{\eta^{\prime}}}{2}, 0, m_{b})
\right)
\nonumber \\
&&-\frac{M}{m_{b}}
\left(
C_{11}(\frac{K}{2}, p_{c}-K,
       \frac{m_{\eta^{\prime}}}{2}, 0, m_{c})
-2C_{12}(\frac{K}{2}, p_{c}-K,
       \frac{m_{\eta^{\prime}}}{2}, 0, m_{c})
+C_{0}(\frac{K}{2}, p_{c}-K,
       \frac{m_{\eta^{\prime}}}{2}, 0, m_{c})
\right)
\nonumber \\
&&-\frac{M(m_{b}-m_{c})}{m_{b}m_{c}}
\left(
C_{11}(\frac{K}{2}-p_{c}, P-K,
       \frac{m_{\eta^{\prime}}}{2}, m_{c}, m_{b})
-2C_{12}(\frac{K}{2}-p_{c}, P-K,
       \frac{m_{\eta^{\prime}}}{2}, m_{c}, m_{b})
\right. \nonumber \\
&&\left. +C_{0}(\frac{K}{2}-p_{c}, P-K,
       \frac{m_{\eta^{\prime}}}{2}, m_{c}, m_{b})
\right) \eea } The three points loop functions and their definitions
are as follows \cite{veltman}:
\begin{equation}
C_0;C_{\mu}(p,k,m_1,m_2,m_3)=\frac{1}{i \pi}\int
d^nq\frac{1;q_{\mu}}{(q^2-m_1^2)((q+p)^2-m_2^2)((q+p+k)^2-m_3^2)},
\end{equation}
where $C_\mu=p_\mu C_{11}+ k_\mu C_{12}$. Using the Feynman
parametrization we obtain the explicit forms of $C_0, C_{11}$ and
$C_{12}$ in terms of Feynman parameters as
\begin{eqnarray}&&C_0=\int_0^1\int_0^{1-x}\frac{-1}{L(x,y)}dx dy\nonumber\\
&&C_{11}=\int_0^1\int_0^{1-x}\frac{1-x}{L(x,y)}dx dy,\,\,\,\,\,\,
C_{12}=\int_0^1\int_0^{1-x}\frac{y}{L(x,y)}dx dy,
\end{eqnarray}
where
\begin{equation}
L= m_2^2+(m_1^2-m_2^2) x-p^2x+p^2x^2-m_2^2y+m_3^2y -k^2y+k^2y^2-2 x
y p.k.
\end{equation}

 For the heavy $b$ and $c$
quarks, it is reasonable to neglect the relative momentum of the
quark constituents and their binding energy relative to their
masses. In this nonrelativistic limit, the constituents are on mass
shell and move together with the same velocity. It implies the
following equations to a good accuracy \be
 M(B_{c})=m_{c}+m_{b}, {\hskip
0.5cm} p_{\bar c}=\frac{m_c}{M}P, {\hskip
0.5cm}p_{b}=\frac{m_b}{M}P.
 \ee
Now let us calculate the decay width of the process \Bcet taking
into account the lepton polarization. Four components of spin vector
of lepton $s_{\mu }$ in terms of $\vec{\eta }$, the unit vector
along the $\ell $
 lepton spin in its rest frame are given by
\be
s_{0}=\frac{\vec{p_{\ell }}\cdot \vec{\eta }}{m_{\ell }%
}\, ,   \vec{s}=\vec{\eta }+\frac{s_{0}}{%
E_{\ell }+m_{\ell }}\vec{p_{\ell }}\,. \label{41} \ee In the
$B^{+}_{c}$ rest frame, the partial decay rate is found to be \be
d\Gamma =\frac{1}{(2\pi )^{3}}\frac{1}{8M_{B_c}}\mid M\mid
^{2}dE_{\eta^{\prime} }dE_{\ell }\,, \label{42} \ee  where \be
|{\cal{M}}|^2=A_{0}(x,y)+(A_{L}\vec{e}_{L}+A_{N}\vec{e}_{N}+A_{T}\vec{e}_{T})\cdot
\vec{\eta}\,, \label{43} \ee
 where $\vec{e}_{i}\ (i=L,N,T)$ is the
unit vector along the longitudinal, normal and transversal
components of the lepton polarization, defined as:
\begin{eqnarray}
\vec{e}_{L} &=&{\frac{\vec{p}_{\ell }}{|\vec{p}_{\ell }|}},  \nonumber \\
\vec{e}_{T} &=&{\frac{\vec{p}_{\ell }\times (\vec{q}\times \vec{p}_{\ell })}{|%
\vec{p}_{\ell }\times (\vec{q}\times \vec{p}_{\ell })|}},  \nonumber \\
\vec{e}_{N} &=&{\frac{\vec{q}\times \vec{p}_{\ell }}{|\vec{q}\times \vec{p}%
_{\ell }|}}\,, \label{44}
\end{eqnarray}
respectively. The quantities  $A_{0} $, $A_{L}$, $A_{N}$, $A_{T}$
can be calculated directly and are given by
\begin{eqnarray}
A_0(t,s)&=& 4M^4_B \{-|f_1|^2 [r_{\eta'}  + r_{\ell})+(-1+t)(-1
+t+s)]\nonumber\\&+&|f_2|^2 [r_2 (-1 + r_{\ell})-(1 + r_{\ell}-t)(1
+ r_{\ell}-t-s)]\nonumber\\&-& 2Re[f_1 f^*_2] [r_2- (1 +
r_{\ell}-t)(-1+ s + t)]\}\label{a0}\nonumber\\\\
A_L(t,s)&=& 2M^4_B \{-|f_1|^2 [(-2 + s +
2t)\sqrt{-4r_{\ell}+t^2}+\sqrt{-4r_{\eta'}+s^2}~t
\cos(z)]\\
&+&|f_2|^2 [(-2r_{\eta'} + s
(1+r_{\ell}-t))\sqrt{-4r_{\ell}+t^2}+\sqrt{-4r_{\eta'}+s^2}~t(-1-r_{\ell}+t)
\cos(z)]\nonumber\\
&+& Re[f_1 f^*_2] [(2-2r_{\eta'}+2r_{\ell}-( 2+ s)t
\sqrt{-4r_{\ell}+t^2}+\sqrt{-4r_{\eta'}+s^2}(-2+t)t
\cos(z)]\}\nonumber\\ \nonumber\\
A_N(t,s)&=&-4M^4_B\sqrt{r_{\ell}}\sqrt{-4r_{\ell}+t^2}\sqrt{-4r_{\eta'}+s^2}\sin(z)
Im[f_1 f^*_2]\\ \nonumber\\
A_T(t,s)&=&-4M^4_B\sqrt{r_{\ell}}\sqrt{-4r_{\eta'}+s^2}\sin(z)\{|f_1|^2+|f_2|^2(1+r_{\ell}-t)-Re[f_1
f^*_2](-2+t)\}
\end{eqnarray}
where $ r_{\eta'}=\frac{M^2_B}{m_{\eta'}^2}$ , $
r_{\ell}=\frac{M_B^2}{m_{\ell}^2},\ t=\frac{2 E_{\ell}}{M_B},\
s=\frac{2 E_{\eta'}}{M_B}$ are normalized energies of the lepton and
$\eta'$, respectively.  $\cos(z)$ is given by: \be
\cos(z)=\frac{2(1+r_2+r_{\ell}-s)+(-2+s)t}{\sqrt{(-4r_{\ell}+t^2)(-4r_{\eta'}+s^2)}}
\ee Here the $z$ is a angle between the final lepton($\ell$) and
$\eta'$ particles.\\
 Using  Eq.($\ref{a0}$) we get the following expression for differential decay rate:
\bea
 \frac{d\Gamma(s)}{ds}
=\frac{\Delta (s)}{8(2\pi)^3}C^2 M^3, \eea where \be
C=\frac{8}{3}\alpha^{2}_s f_{B_c}A_{\eta^{\prime}}
\frac{G_F}{\sqrt{2}}V_{cb}. \ee
 and the expression for $\Delta$ is  as:
 \bea \Delta &=& \frac{(1+r_{\eta'}-r_{\ell}-s)^2
 \sqrt{-4r_{\eta'}+s^2}}{3(1+r_{\eta'}-s)^3} \Big\{2(1+r_{\eta'}-s)(-4r_{\eta'}+s^2)
 (|f_1|^2+|f_2|^2+2Re[f_1 f^*_2 ])\nonumber \\&-& 4r_{\ell}\{|f_1|^2 (-3+ r_{\eta'}+3s-s^2)
 +|f_2|^2(r_{\eta'}-3r^2_{\eta'}+3r_{\eta'}s-s^2)\nonumber\\&+& Re[f_1 f^*_2](8r_{\eta'}-3-3r_{\eta'} s + s^2)\}
 \Big\}
 \eea
 It  should be mentioned that if one neglects the lepton
 mass($r_{\ell}=0$), the results in \cite{akio}  are obtained.
If we define the longitudinal, normal and transversal $\ell$
polarization asymmetries by \be P_{i}(s)={\frac{d\Gamma
(\vec{e}_{i})-d\Gamma (-\vec{e}_{i})}{d\Gamma ( \vec{e}_{i})+d\Gamma
(-\vec{e}_{i})}}\,,\ (i=L,N,T)\,, \label{49} \ee
 we find that
\be \label{Pi} P_{i}(s)={\frac{\int A_{i}(t,s)dt}{\int
A_{0}(t,s)dt}}\,,\ (i=L,N,T)\,. \label{410} \ee
\section{Numerical Results and Discussions}
In this section the numerical  analysis is done not only for the
differential decay width but also for the polarization
asymmetries($P_i$). For numerical results, we take $\alpha_s
=\alpha_s (M_{B_c})=0.2$,
 $V_{cb}=0.04$, $A_{\eta^{\prime}}=0.2$ and $\tau_{B_c} =0.46ps$.
The decay constant $f_{B_c}$ probes the strong(nonpertubative) QCD
dynamics which bind $b$ and $\bar c$  quarks to form the bound state
$B_c$.  In nonrelativistic limit, $f_{B_c}$ can be related to the
value of  the $B_c$ wave function at origin\cite{van}. Leptonic
decay constant is estimated by QCD sum rules\cite{aliev} and using
the nonrelativistic potential models\be
f_{B_c}=\left\{\begin{array}{ll}
       450 MeV & \mbox{(Buchm{\"u}ller-Tye potential\cite{p1})}\\
       512 MeV & \mbox{(power law potential\cite{p2})}\\
       479 MeV & \mbox{(logarithmic potential\cite{p3})}\\
       687 MeV & \mbox{(cornell potential\cite{p4})}
\end{array}\right.
\ee
 For numerical illustrations, we take $f_{B_c}=0.5\, Gev$.
We also do numerical integration in eq. (\ref{Pi}) for $s$ values in
the interval $s\epsilon[2 \sqrt{r_{\eta'}},1+ r_{\eta'}-r_{\ell}]$
with  respect to $t$ ranging from $t_{Min} $ to $t_{Max} $  which
are given as: \bea t_{Min}&=& \frac{(1 + r_{\eta'} + r_{\ell} - s)(2
- s)}{2(1+r_{\eta'}  - s)}- \frac{|(1 + r_{\eta'} - r_{\ell} -
s)|\sqrt{(-4 r_{\eta'} + s^2)}}{2(1 + r_{\eta'} - s)}\nonumber\\
t_{Max}&=& \frac{(1 + r_{\eta'} + r_{\ell} - s)(2 -
s)}{2(1+r_{\eta'} - s)}+ \frac{|(1 + r_{\eta'} - r_{\ell} -
s)|\sqrt{(-4 r_{\eta'} + s^2)}}{2(1 + r_{\eta'} - s)} \eea
 The dependence of the the branching ratio on normalized $\eta^{\prime}$ momentum
  for $ \mu$ and $\tau$ cases are displayed in
Fig.1 and Fig. 2.  It is seen that the normalized $\eta^{\prime}$
momentum distribution is peaked at small values of $s$. In fact, it
is reasonable if we consider the expressions of $f_1$ and $f_2$ in
terms of basic scalar functions $C_0$ and $C_{12}$ and $C_{11}$ in
\cite{veltman}, the normalized $\eta^{\prime}$ momentum distribution
 behaves as \be {\propto}\frac{1}{ \sqrt{s^{2}-r_{\eta^{\prime}} }},
\ee when $s$ is small. Therefore, there is a  singularity at the
starting point of the distribution, but it is integrable and give
finite decay width. The branching ratio is estimated to be \be
Br(B_c \to \eta' \mu \bar{\nu}~)\sim 1.6\times 10^{-4}, \ee \be
Br(B_c \to \eta' \tau \bar{\nu}~)\sim 2.2\times 10^{-4}, \ee for
$\mu$ , $e$ leptons and
 $\tau$ leptons, respectively.

 Fig.3 and Fig.4 are displaying the dependency of $P_L$ for $\mu$
 and $\tau$ leptons, respectively. We see that the $P_L$ for $\tau$
 lepton can take both negative and positive values. Precisely, for
 $s\leq 0.45$ it takes negative values and elsewhere it is
 positive.

 Fig.5 and Fig.6 are displaying the dependency of $P_N$ for $\mu$
 and $\tau$ leptons, respectively. We see that for both leptons $P_N$
 is negative and has minimum  at $s\simeq 0.83$ and $s\simeq 0.75$ , respectively.

 Fig.7 and Fig.8 are displaying the dependency of $P_T$ for $\mu$
 and $\tau$ leptons, respectively. We see that for both leptons $P_T$
 is negative and has minimum  at $s\simeq 0.95$ and $s\simeq 0.75$ , respectively.

 Finally, a few words about the detectibilty of the lepton polarization asymmetries at $B$ factories or future
hadron colliders, are in order. As an estimation, we choose the
averaged values of the longitudinal, transversal and normal
polarizations for both $\mu$ and $\tau$ leptons (see TABLE 1).
\begin{center}
TABLE 1. The averaged Longitudinal, Normal and Transversal
polarization for $\mu$ and $\tau$ leptons. \vskip 0.5cm

\begin{tabular}{|c|c|c|}
  \hline
  $\langle P_i\rangle$&$B_c\rightarrow \eta' \mu \nu_\mu$ & $B_c\rightarrow \eta' \tau \nu_\tau$ \\\hline
  $\langle P_L\rangle$ & 0.71 & 0.258 \\
 $\langle P_N\rangle$ & -0.007 & -0.1 \\
  $\langle P_T\rangle$& -0.009 & -0.09 \\
  \hline
\end{tabular}
\end{center}
Experimentally, to measure an asymmetry $\langle P_i\rangle$ of a
decay with the branching ratio $B$ at the $n\sigma$ level, the
required number of events is given by the formula $N=n^2/({\cal{B}}
\langle P_i\rangle^2 )$. It follows from this expression and TABLE 1
that to observe the lepton polarizations  $\langle P_L\rangle\,,
\langle P_N\rangle$ and $\langle P_T\rangle$ in \Bcet decay at
$1\sigma$ level, the expected number of events are
$N=(1,3,10^4)\times10^7$, respectively. On the other hand, the
number of $B\overline{B}$ pairs that are expected to be produced at
B factories and LHCb will be $10^8$ and $10^{12} B\overline{B}$
pairs, respectively. A comparison of these numbers allows us to
conclude that not only the measurements of the longitudinal
polarization of muon and longitudinal, normal and transversal
polarization of $\tau$ lepton, but also the measurements of the
normal and transversal polarizations of the $\mu$ lepton with the
order of $\approx \% 1$ (see TABLE 1) could be accessible at $B$
factories.

 In conclusion, we carried out
a study on the semileptonic
annihilation decays \Bcet and lepton polarization asymmetries.\\
\bigskip
\noindent {\large\bf Acknowledgment}

 The author thanks TM Aliev for
helpful discussions.

\newpage
\begin{center}
\large{Figure Captions}
\end{center}
\vspace{4cm} Figure 1: The distribution of $dBr/ds$ as a function of
$s$(normalized energy of $\eta'$) for the $B_c\rightarrow \eta'
\mu\nu_{\mu}$ decay.\\
Figure 2: The same as Figure 1 but for  for the $B_c\rightarrow
\eta'
\tau\nu_{\tau}$ decay.\\
Figure 3:The dependence of the longitudinal lepton polarization
$P_L$ on $s$  for the $B_c\rightarrow \eta' \mu\nu_{\mu}$ decay.\\
Figure 4: The same as Figure 3 but for the $B_c\rightarrow \eta'
\tau\nu_{\tau}$ decay.\\
Figure 5:The dependence of the normal lepton polarization $P_N$ on
$s$ for the $B_c\rightarrow \eta' \mu\nu_{\mu}$ decay.\\
Figure 6: The same as Figure 5 but for the $B_c\rightarrow \eta'
\tau\nu_{\tau}$ decay.\\
Figure 7:The dependence of the transversal lepton polarization $P_T$
on $s$ for the $B_c\rightarrow \eta' \mu\nu_{\mu}$decay.\\
Figure 8: The same as Figure 7 but for the $B_c\rightarrow \eta'
\tau\nu_{\tau}$decay.\\
\newpage
\begin{center}
\begin{figure}
\vskip 3.5 cm
    \includegraphics{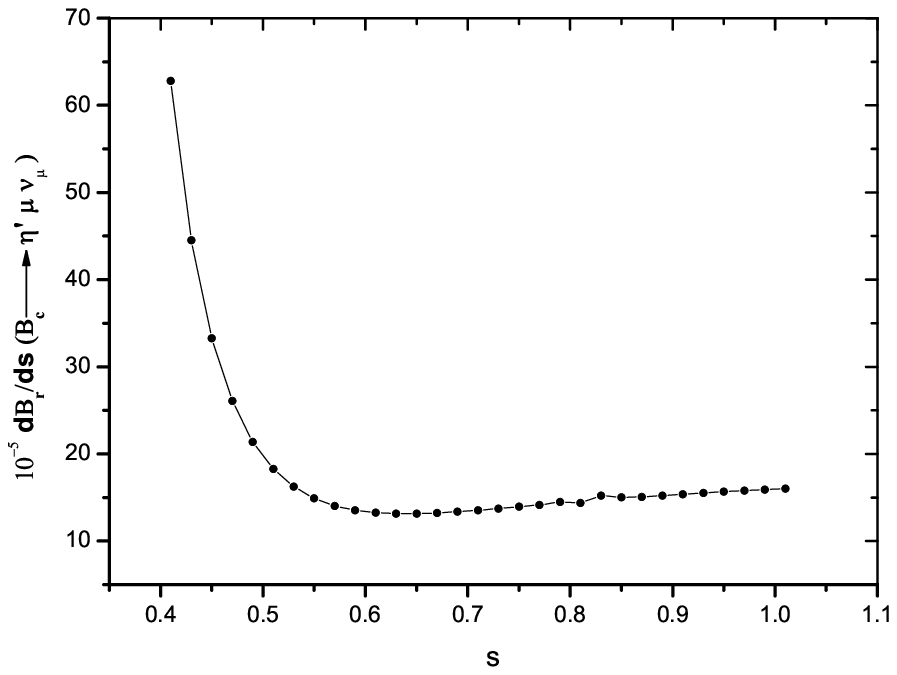}
\vskip 2.8cm \caption{}\label{brmu1}
\end{figure}
\end{center}
\begin{center}
\begin{figure}
\vskip 3.70 cm
    \includegraphics{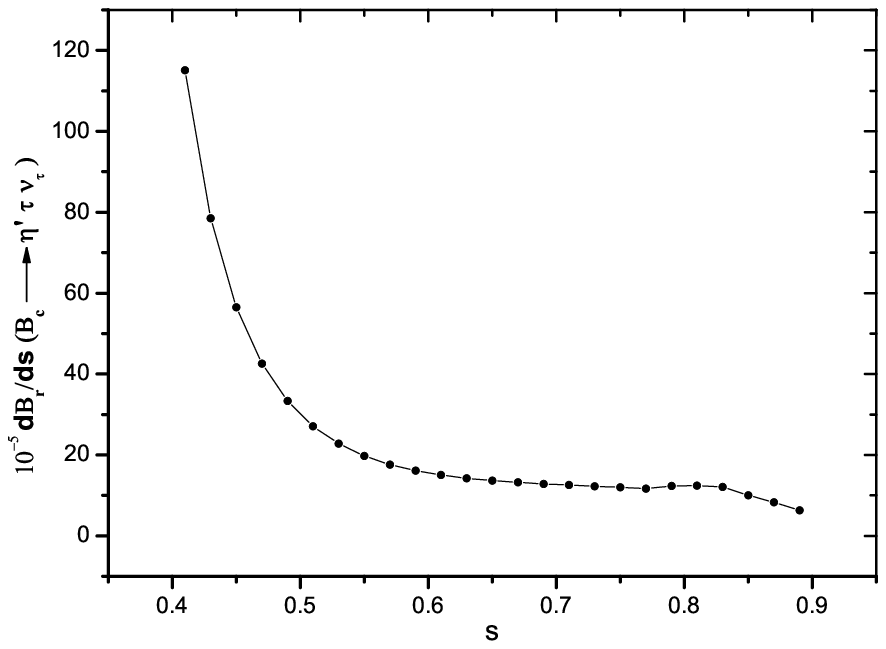}
\vskip 2.8cm \caption{}\label{brmu2}
\end{figure}
\end{center}
\begin{center}
\begin{figure}
\vskip -1.9 cm
    \includegraphics{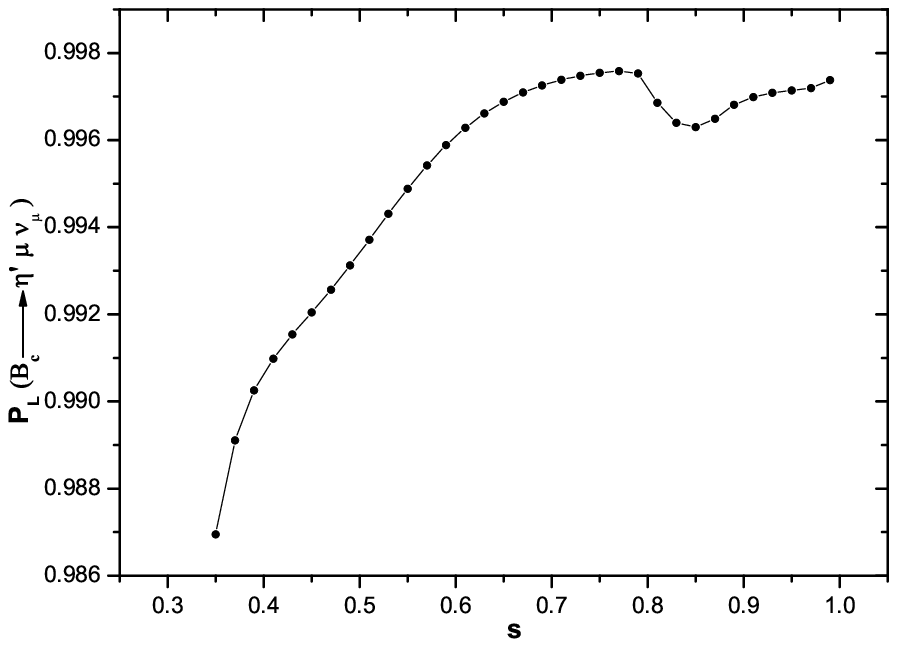}
\vskip 2.8cm \caption{}\label{brmu3}
\end{figure}
\end{center}
\newpage
\begin{center}
\begin{figure}
\vskip 3.5 cm
    \includegraphics{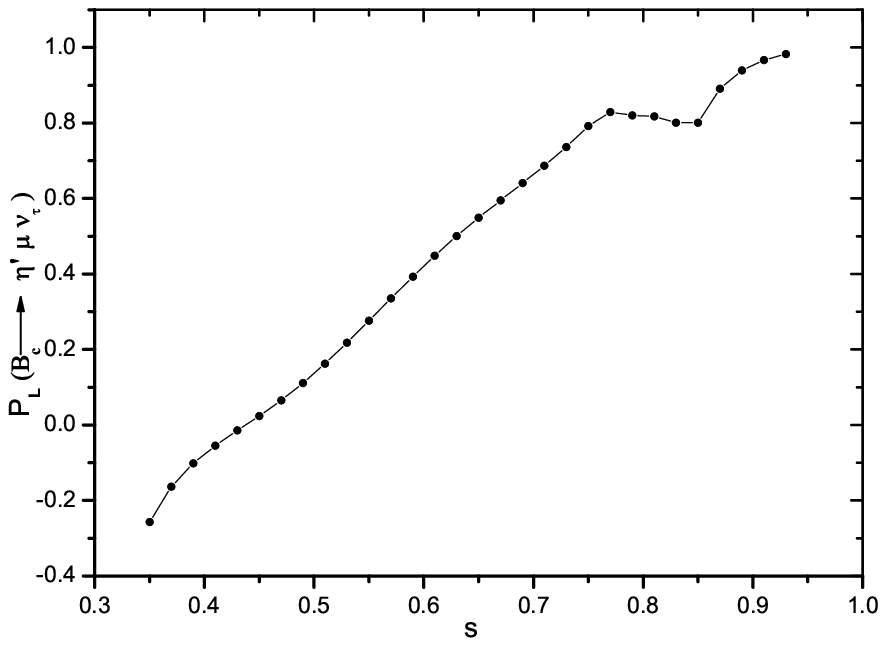}
\vskip 2.8cm \caption{}\label{brmu4}
\end{figure}
\end{center}
\begin{center}
\begin{figure}
\vskip 3.70 cm
    \includegraphics{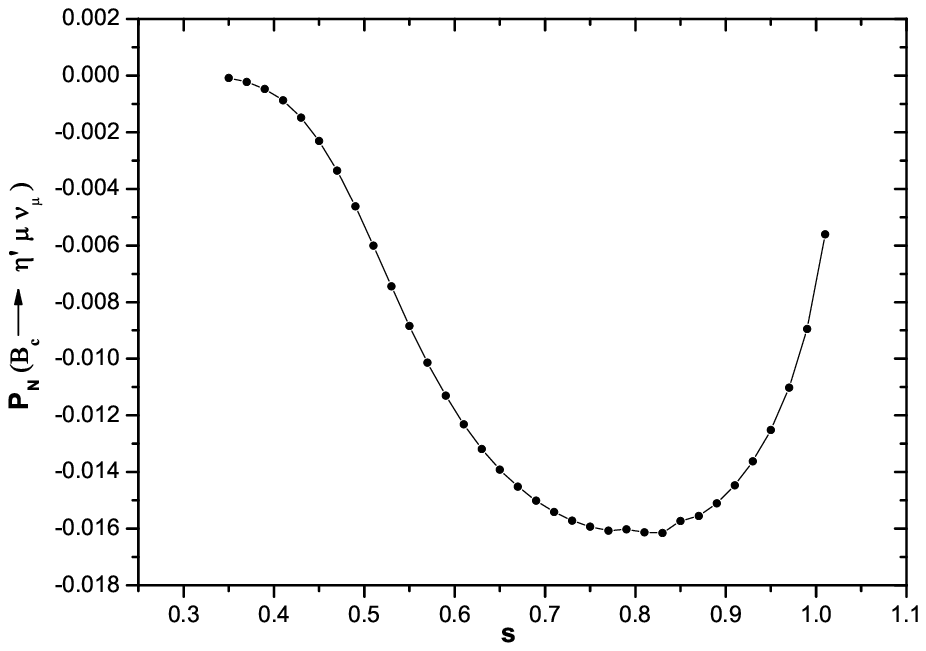}
\vskip 2.8cm \caption{}\label{brmu5}
\end{figure}
\end{center}
\begin{center}
\begin{figure}
\vskip -1.9 cm
    \includegraphics{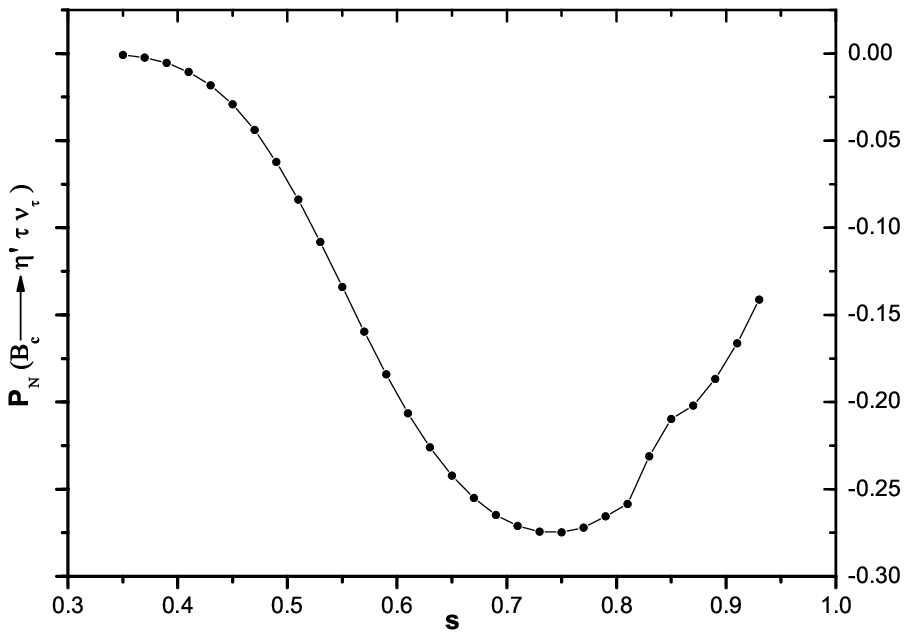}
\vskip 2.8cm \caption{}\label{brmu6}
\end{figure}
\end{center}
\newpage
\begin{center}
\begin{figure}
\vskip 3.5 cm
    \includegraphics{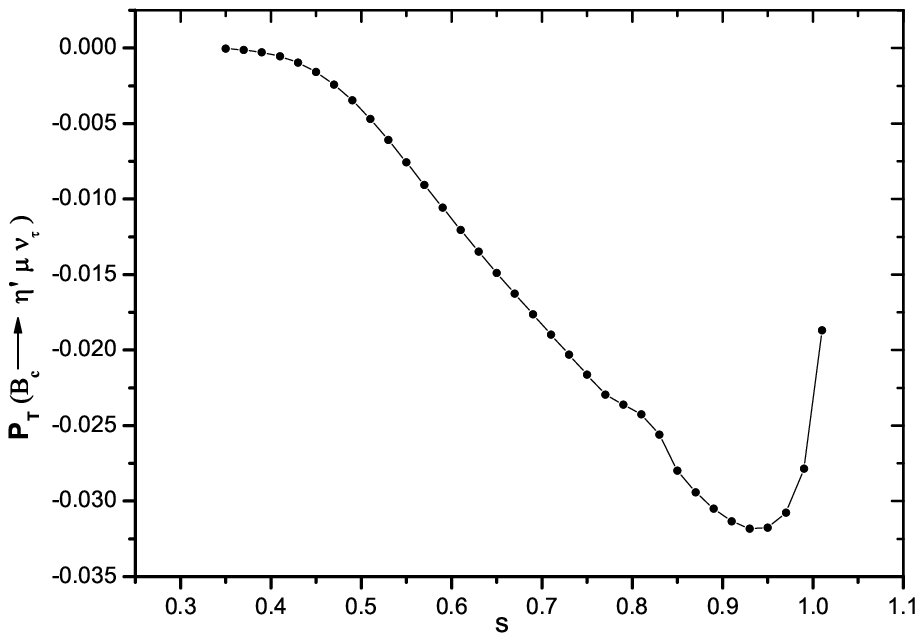}
\vskip 2.8cm \caption{}\label{brmu7}
\end{figure}
\end{center}
\begin{center}
\begin{figure}
\vskip 3.70 cm
    \includegraphics{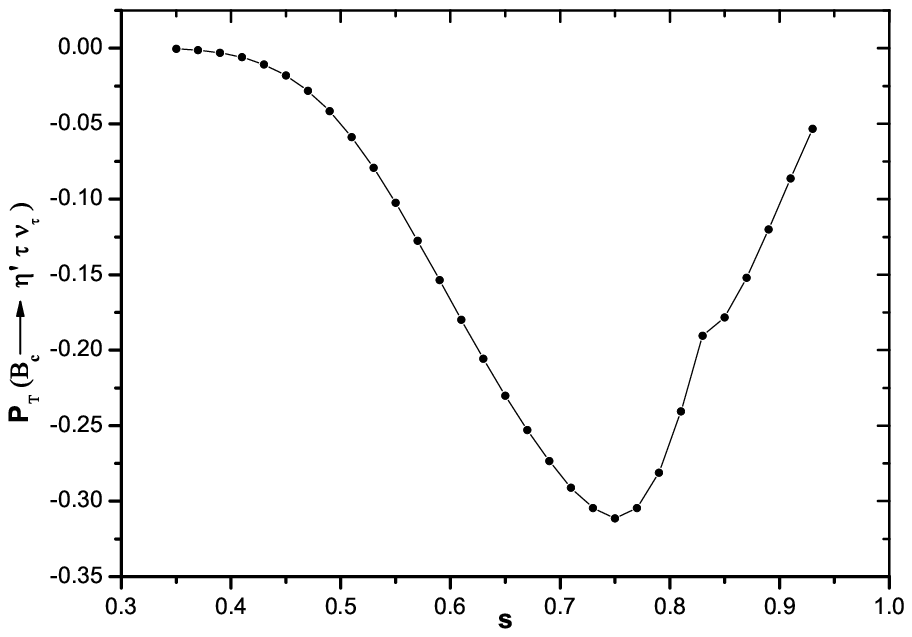}
\vskip 2.8cm \caption{}\label{brmu8}
\end{figure}
\end{center}
\end{titlepage}
\end{document}